\documentclass[aps,twocolumn,prb,floatfix,showpacs,superscriptaddress]{revtex4}

\usepackage{amsmath}
\usepackage{amssymb}
\usepackage{amsfonts}
\usepackage[pdftex]{graphicx}
\usepackage{units}
\usepackage{bm}
\usepackage{color}
\usepackage{tikz}
\usepackage{color}
\usepackage{array}

\definecolor{dunkelgrau}{rgb}{0.8,0.8,0.8}
\definecolor{hellgrau}{rgb}{0.95,0.95,0.95}

\newcommand{\amend}[1]{#1}
\begin{document}
\title{Thermal String Excitations in Artificial Spin-Ice Dipolar Arrays}

\author{Danny Thonig}
\affiliation{Max-Planck-Institut f\"ur Mikrostrukturphysik, D-06120 Halle (Saale), Germany}
\affiliation{Institut f\"ur Physik, Martin-Luther-Universit\"at Halle-Wittenberg, D-06099 Halle (Saale), Germany}
\email{dthonig@mpi-halle.mpg.de}

\author{Stephan Rei\ss{}aus}
\affiliation{Institut f\"ur Physik, Martin-Luther-Universit\"at Halle-Wittenberg, D-06099 Halle (Saale), Germany}

\author{Ingrid Mertig}
\affiliation{Max-Planck-Institut f\"ur Mikrostrukturphysik, D-06120 Halle (Saale), Germany}
\affiliation{Institut f\"ur Physik, Martin-Luther-Universit\"at Halle-Wittenberg, D-06099 Halle (Saale), Germany}

\author{J\"urgen Henk}
\affiliation{Institut f\"ur Physik, Martin-Luther-Universit\"at Halle-Wittenberg, D-06099 Halle (Saale), Germany}

\date{\today}

\begin{abstract}
 We report on a theoretical investigation of artificial spin-ice dipolar arrays, using a \amend{nanoisland} geometry adopted from recent experiments [A. Farhan \textit{et al.}, Nature Phys.\ \textbf{9} (2013) 375]. The number of thermal magnetic string excitations in the square lattice is drastically increased by a vertical displacement of rows and columns. We find large increments especially for low temperatures and for string excitations with quasi-monopoles of charges $\pm 4$. By kinetic Monte Carlo simulations we address the thermal stability of such excitations, thereby providing time scales for their experimental observation.
\end{abstract}









\maketitle

\section{Introduction}
Frustrated magnetic systems have become a topic of particular interest in condensed matter physics\cite{Schiffer02,Hodges11,Hamann12}. The geometrical frustration arises from the specific geometry of the system, rather than from disorder. It leads to `exotic' low-temperature states, for example spin ice. In pyrochlore lattices---prominent compounds are dysprosium and holmium titanate---, the spins arranged in corner-sharing tetrahedra mimic the hydrogen positions in water ice\cite{Harris97}. Experiments have found evidence for the existence of magnetic monopoles in these materials\cite{Bramwell01,Gingras09}, showing properties of hypothetical magnetic monopoles postulated to exist in vacuum\cite{Dirac31}. But also nano-scale arrays of ferromagnetic single-domain islands can show an artificial spin ice\cite{Wang06,Harris07a}.

Artificial spin ice consists of twodimensional periodic arrangements of nanometer-sized magnets. These nanoislands are typically elongated to show a single-domain state\cite{DeBell97,Stamps99}, modeled for example as a magnetic dipole; the magnetic moment of a single island then points in one of two directions. Because the nanoislands are isolated from each other---e.\,g., separated by a distance of the order of several hundred nanometer---, they are coupled by the long-range dipole-dipole interaction\cite{Remhof08,Mengotti09}.

\begin{figure}
 \centering
 \includegraphics[width = 0.9\columnwidth]{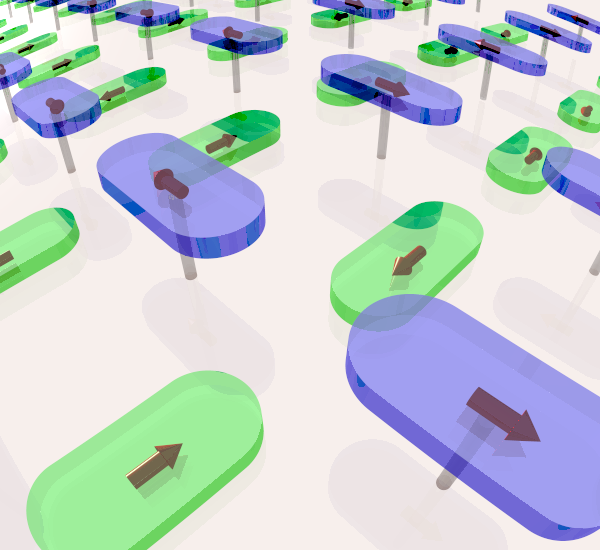}
 \caption{(Color online) Artist's rendering of artificial spin ice on a square lattice of nanomagnets with vertically displaced rows and columns. Arrows indicate the direction of the magnetic moments.}
 \label{fig:lattice}
\end{figure}

Typical geometries of the nano-scale arrays are honeycomb or square lattices, fabricated using microstructuring techniques which allow for fine-tuning to obtain specific properties\cite{Rougemaille11}. Shifting the rows and columns of a square lattice vertically  (Fig.~\ref{fig:lattice}), by an amount determined by the lattice spacing and the islands' dimensions, one can produce the same degree of degeneracy in the ground state as in pyrochlore spin ice\cite{Moeller06,Mol10} \amend{and the same residual entropy as water ice at zero temperature \cite{Pauling35}}. 

Because of the specific geometries used so far, artificial spin ice was hardly thermally active\cite{Wysin13}: for permalloy nanomagnets, the magnetic moment of each island is in the order of some $10^{7}$ Bohr magneton, equivalent to an interaction energy of about $\unit[10^{-19}]{J}$. Thus in simulations, the activation temperature is much larger than the melting temperature of permalloy ($\unit[1450]{K}$). However in recent investigations by Farhan and coworkers, thermal activation at \amend{$T = \unit[420]{K}$} has been shown for up to three hexagons of nanomagnets\cite{Farhan13} and for a square spin-ice lattice\cite{Farhan13a}. \amend{The theoretical investigation presented in this paper relies on the experimentally feasible nanoisland dimensions of Ref.~\onlinecite{Farhan13} in order to study thermal excitations at room temperature for square spin ice lattices.}  As one result, we confirm the thermal activation at \amend{$T = \unit[420]{K}$ and below (e.\,g., at room temperature)} found experimentally. Moreover, the calculated switching rates of the nanomagnets are in the order of $\unit[0.2]{s^{-1}}$, thus accessible by experimental techniques.

In a ground state, the square-lattice nanomagnets align according to the ice rule (`two in, two out')\cite{Mol09,Mol10}. Associating a magnetic charge $Q$ to each node of the square lattice, a ground state is characterized by $Q = 0$ at each node. Excitations appear as reversals of dipoles, leading to nodes with a charge of $\pm 2$ or $\pm 4$. String excitations\cite{Mol09} are then given by a pair of these emergent quasi-monopoles\cite{Mengotti10,Hugli12} with opposite charge that are connected by a ferromagnetic path of nanomagnets\cite{Mengotti11,Pushp13,Farhan13a}. While these strings have been produced by experimentally by an external magnetic field, we focus in this paper on their thermal excitation. The response of the system to an external perturbation is observed by a variety of experimental techniques, for example photoemission electron microscopy\cite{Mengotti08,Farhan13a}.

In the most part, string excitations with $|Q| = 2$ nodes have been considered so far\cite{Moeller09,Mol09,Mol10}, which is attributed to the comparably small probability of $|Q| = 4$ string excitations. We show in this paper that the above-mentioned vertical displacement in the square lattice leads to a drastic increase of the number of $|Q| = 4$ string excitations, in particular at low temperatures. Moreover, we address the thermal stability of such excitations, thereby providing time scales for their experimental observation.

The paper is organized as follows. The theory is outlined in Section~\ref{sec:theory}, results are discussed in Section~\ref{sec:discussion}. Appendices comprise information on the dipolar energies (\ref{sec:energies}) and the Monte Carlo simulations (\ref{sec:Monte}).

\section{Theoretical aspects}
\label{sec:theory}
In this Section we address those aspects of the theoretical approach needed for the discussion of the results. For more details, we refer to the appendices.

For the present study we consider nanomagnets with dimensions taken from Ref.~\onlinecite{Farhan13} (length $\unit[470]{nm}$, width $\unit[170]{nm}$, and height $\unit[3]{nm}$)\amend{, since these exhibit thermal excitations at experimentally achievable temperatures}. Each nanomagnet in the sample is labeled by an index $i$. The lattice constant $a$ of the square lattice\cite{Wang06} is $\unit[793.8]{nm}$ (the lattice spacing in Ref.~\onlinecite{Farhan13a} is $\unit[425]{nm}$). Due to their elongated shape and magnetic anisotropy, they are in a single-domain state with magnetization parallel to the long edges of the islands. Their magnetic state is thus well described by a magnetization vector $\pm \vec{M}_{i}$.  For permalloy islands of the above size one has $|\vec{M}_{i}| \approx \unit[200 \cdot 10^{3}]{A m^{-1}}$. Rows and columns are vertically displaced by $\delta z$, given in units of $a$. Strictly speaking, the twodimensional lattice is turned quasi-twodimensional for $\delta z \not= 0$ (Fig.~\ref{fig:lattice}).

Instead approximating the nanomagnets as points\cite{Moeller06,Mol09,Mol10} or dipolar needles\cite{Moeller06,Moeller09}, we compute the dipole-dipole energies for realistic shapes. The computation of the dipole-dipole energies is done numerically, allowing in principle for arbitrarily shaped nanoislands. It turns out that the dipolar interaction\cite{Mengotti08,Mengotti09} is relevant only for first-nearest neighbors and for second-nearest neighbors, with energies $E_{\mathrm{1NN}}$ and $E_{\mathrm{2NN}}$, respectively\cite{Vedmedenko05}.

\begin{figure}
 \centering
 \includegraphics[scale=1.0]{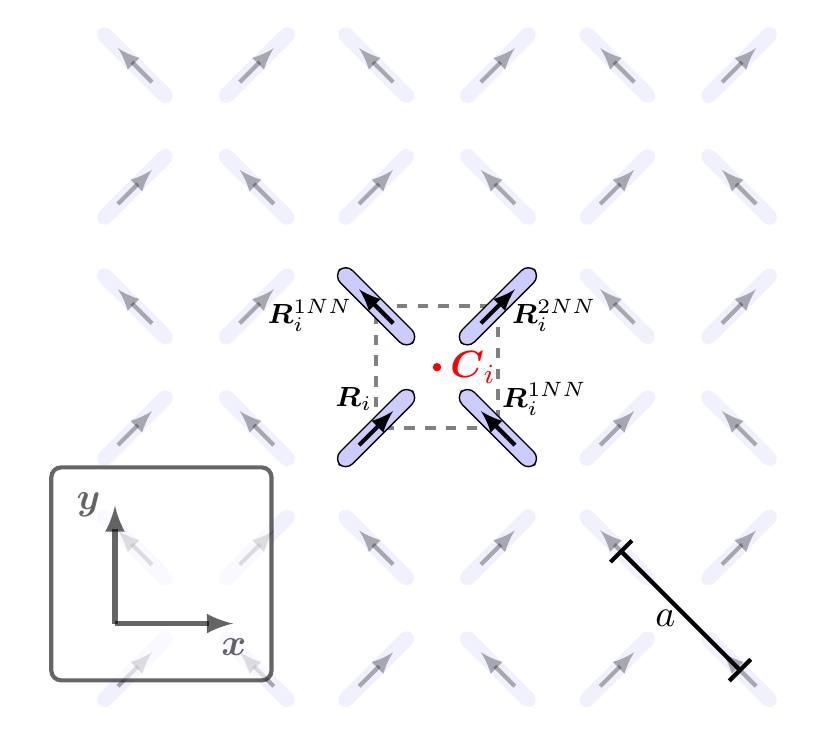}
 \caption{(Color online) Square lattice of nanomagnets, forming a dipolar array. A node with center $\vec{C}_{i}$ is indicated by the dashed square and shows a `2in2outAd' configuration. Magnetic moments $\vec{M}_{i}$ \amend{of islands $R_i$ and the two first nearest neighbors of $R_i$ are represented as arrows and by $R_i^{1NN},R_i^{2NN}$, respectively.} \amend{The lattice parameter $a$ is illustrated by the bold bar.} The inset displays the Cartesian axes.}
 \label{fig:square}
\end{figure}

The center $\vec{C}_{i}$ of a node $i$ that consists of four nanomagnets at positions $R_{j}$ (Fig.~\ref{fig:square}),
\begin{align}
 \vec{C}_{i} & = \frac{1}{4} \sum_{N_{i}} \vec{R}_{j} = \vec{R}_{i} + \frac{a}{2} (\vec{e}_{x} + \vec{e}_{y}),
\end{align}
carries a charge $Q_{i}$. This charge is defined by the number of magnetic dipoles pointing toward this node,
\begin{align}
 Q_{i} & \equiv \sum_{j \in N_{i}} \frac{\vec{M}_{j} \cdot (\vec{C}_{i} - \vec{R}_{j})}{|\vec{M}_{j}| |\vec{C}_{i} - \vec{R}_{j}|}
 \label{eq:charge}
\end{align}
leading to $Q_{i} \in \{0, \pm 2, \pm 4 \}$.

The different magnetic configurations of the nodes are defined in Table~\ref{tab:configurations}. The ice rule predicts groundstate configurations `2in2out' (Ref.~\onlinecite{Morgan10}) which appear in two flavors: `2in2outAd' shows inward pointing moments at adjacent (`Ad') nanomagnets, whereas `2in2outOp' shows inward pointing moments at opposite (`Op') nanomagnets.

\begin{table}
 \caption{Magnetic configurations of nodes, defined in Ref.~\onlinecite{Morgan10}. Charges are defined in eq.~(\ref{eq:charge}). The multiplicity gives the degree of degeneracy for each configuration. The energy of a node is expressed in terms of the first- and second-nearest neighbor energies $E_{\mathrm{1NN}}$ and $E_{\mathrm{2NN}}$.}
 \label{tab:configurations}
 \centering
 \begin{tabular}{lm{1cm}rcc}
  \hline \hline
  Configuration & &\multicolumn{1}{c}{charge} & \multicolumn{1}{c}{multiplicity} & \multicolumn{1}{c}{energy}
  \\
  \hline
  \rule{0pt}{25pt} `4in'         & \includegraphics[scale=0.5]{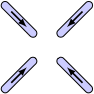} & $+4$   & 1 & $4 E_{\mathrm{1NN}} + 2 E_{\mathrm{2NN}}$ \\
  \rule{0pt}{25pt} `3in1out'     & \includegraphics[scale=0.5]{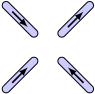} & $+2$   & 4 & $0$ \\
  \rule{0pt}{25pt} `2in2outAd'   & \includegraphics[scale=0.5]{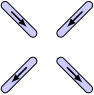} & $0$    & 4 &  $- 2 E_{\mathrm{2NN}}$\\
  \rule{0pt}{25pt} `2in2outOp'   & \includegraphics[scale=0.5]{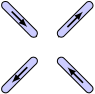} & $0$    & 2 &  $- 4 E_{\mathrm{1NN}} + 2 E_{\mathrm{2NN}}$ \\
  \rule{0pt}{25pt} `1in3out'     & \includegraphics[scale=0.5]{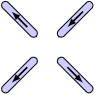} & $-2$   & 4 & $0$ \\
  \rule{0pt}{25pt} `4out'        & \includegraphics[scale=0.5]{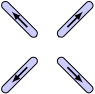}  & $-4$   & 1 & $4 E_{\mathrm{1NN}} + 2 E_{\mathrm{2NN}}$ \\
  \hline \hline
 \end{tabular}
\end{table}

\begin{figure}
 \centering
 \includegraphics[width = 0.9\columnwidth]{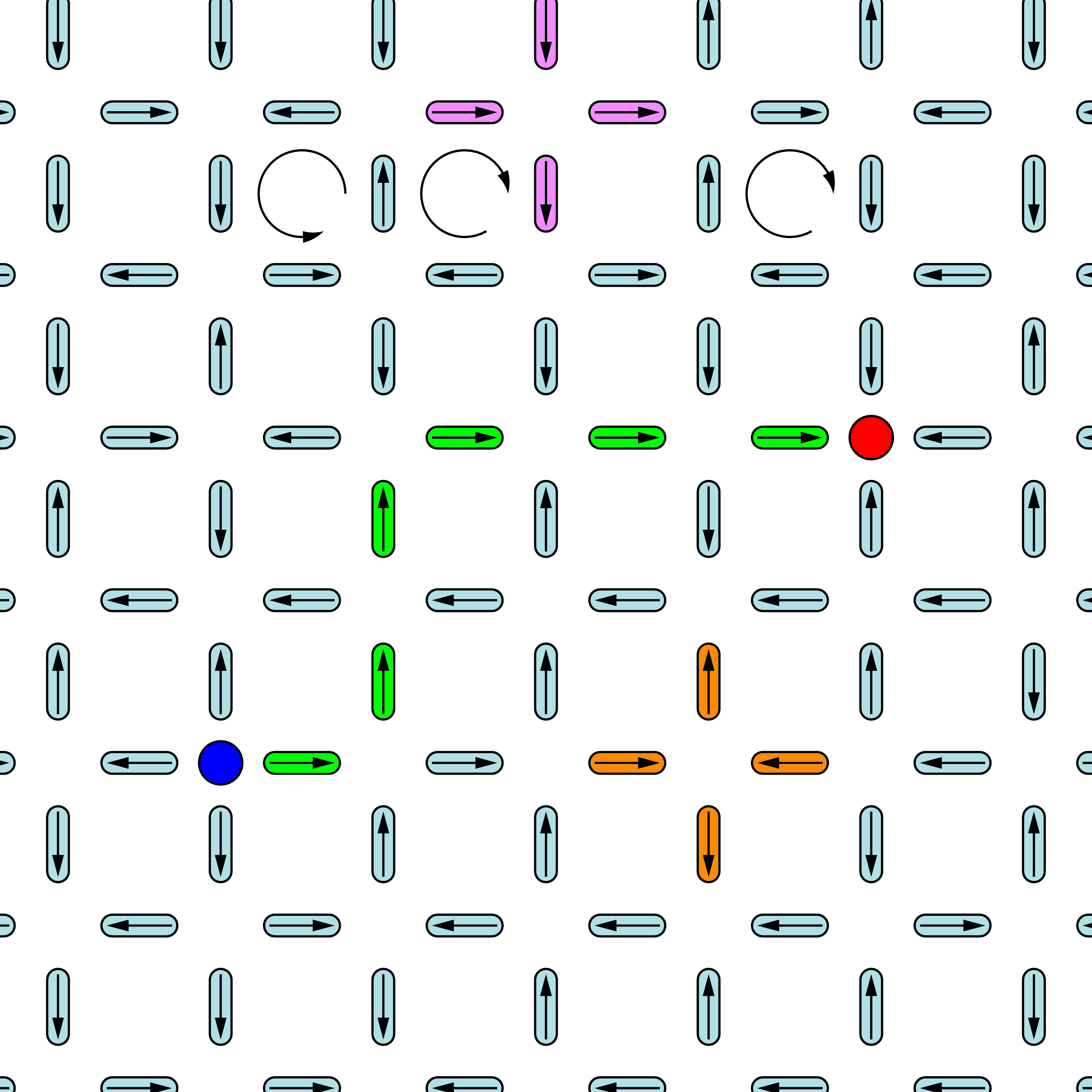}
 \caption{(Color) Snapshot of a magnetic configuration in square-lattice spin ice with $\delta z = 0.27\,a$ ($a$ lattice constant). A string excitation is formed by a quasi-monopole with charge $Q = -4$ (node with the blue circle) connected by a ferromagnetic string (path of six green nanomagnets) with a quasi-monopole with $Q = +4$ (node with the red circle). Arrows in each nanomagnet indicate the respective magnetization orientation. The four purple (orange) nanomagnets form a node with `2in2outAd' (`2in2OutOp') configuration. For three plaquettes, the orientation of flux closures are shown by circular arrows. The snapshot, taken from a kinetic Monte Carlo simulation at $T = \unit[300]{K}$, shows a part of the entire sample.}
 \label{fig:string}
\end{figure}

In accordance with  the point group symmetry of the nodes, the configurations are degenerate, as given by their multiplicity (Table~\ref{tab:configurations}). For $\delta z = 0$, the least energy is produced by nodes with a `2in2outOp' arrangement ($-4 E_{\mathrm{1NN}} + 2 E_{\mathrm{2NN}}$), with multiplicity $2$ (see the four orange nanomagnets in Fig.~\ref{fig:string}; cf.\ also Ref.~\onlinecite{Mol10}). The `2in2outAd' configuration (confer the four purple nanomagnets in Fig.~\ref{fig:string}) has an energy of $E = -2 E_{\mathrm{2NN}}$ and a multiplicity of $4$.

\begin{figure}
 \centering
 \includegraphics[width = 0.9\columnwidth]{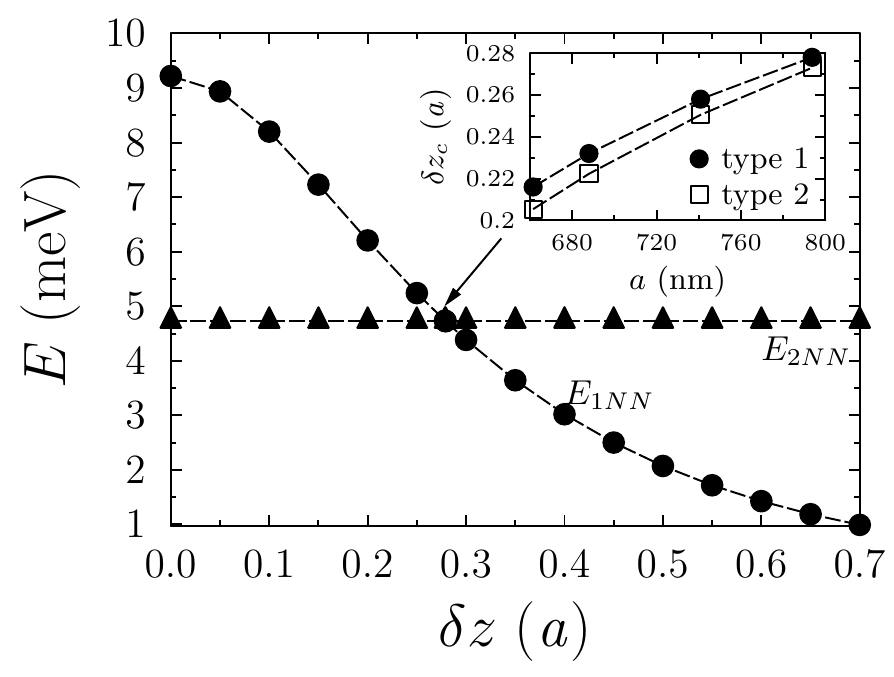}
 \caption{Dipolar energies of nanomagnets on a square lattice. Energies of first-nearest (1NN, filled circles) and second-nearest (2NN, filled triangles) nanoislands are shown versus the vertical displacement $\delta z$ (in units of the lattice constant $a$). The arrow marks $\delta z = 0.27\,a$, for which $E_{\mathrm{1NN}} =  E_{\mathrm{2NN}}$. The inset displays this critical point as a function of the lattice constant $a$. \amend{Here, filled circles and open squares indicate the crossing for rectangular (type 1) and rounded (type 2) islands, respectively.} The dimensions of the nanomagnets and the lattice parameters are given in Section~\ref{sec:theory}.}
 \label{fig:energies}
\end{figure}

An increasing vertical displacement $\delta z$ of rows and columns in the lattice results in a decrease of $E_{\mathrm{1NN}}$ (Fig.~\ref{fig:energies}). $E_{\mathrm{2NN}}$ is unchanged because second-nearest neighbors are on the same or on adjacent rows or columns. At the special $\delta z$ for which $E_{\mathrm{1NN}} = E_{\mathrm{2NN}}$ the degeneracy of the nodes' ground state is increased to $4+2=6$ (Ref.~\onlinecite{Mol10}). The honeycomb lattice possesses the same degree of degeneracy: the frustrated least-energy nodes with charges $\pm 1$ (`2in1out' or `2out1in') have a multiplicity of $3$ each; showing  identical energies, they are $6$-fold degenerate\cite{Mengotti10}. \amend{It is important to mention that in the honeycomb lattice this sixfold degeneracy is out of 8 possible vertices; in the square lattice considered here the sixfold degeneracy is out of 16 vertices (cf. Table~\ref{tab:configurations}). However, one may consider both lattices and their magnetic ground states equivalent because both have the same residual entropy of $0.2\, k_{\mathrm{B}}$ (Appendix \ref{sec:entropy}). Furthermore, the approach of $E_{\mathrm{1NN}}$ to $E_{\mathrm{2NN}}$ reduces the total energy and, thus, enhances the thermal activity, allowing simulations already for room temperature.}

For the present samples, we obtain $\delta z = 0.27\,a$, which is a monotonous function of the lattice constant $a$ (inset in Fig.~\ref{fig:energies}). This value differs from those calculated for nano-scale arrays consisting of point or dipolar needles ($0.419\,a$ in Ref.~\onlinecite{Moeller06} and $0.444\,a$ in Ref.~\onlinecite{Mol10}).  

\amend{The $\delta z$ for which $E_{\mathrm{1NN}} = E_{\mathrm{2NN}}$ depends also moderately on the island shape (inset in Fig.~\ref{fig:energies}). To check this we studied rectangular islands (type 1) with an aspect ratio of $2.76$ (as in Ref.~\onlinecite{Farhan13}) and rounded islands (type 2). The latter have the same area as type-1 islands but are composed of a rectangle with an aspect ratio of $1.98$ and two terminating semi-circles with radii of $\unit[85]{nm}$. The results presented in this Paper are for islands of type 2.}

A string excitations is identified as a ferromagnetic path of nanoislands connecting a pair of nodes with opposite nonzero charges (Fig.~\ref{fig:string}). To quantify the thermal activation, we address the fraction of nodes with charge $Q$  in the sample, $\eta_{Q} \equiv N_{Q} / N$; on average $\langle \eta_{Q} \rangle = \langle \eta_{-Q} \rangle$. 

In this paper, we report on results for a lattice with $20 \times 20$ cells with $2$ nanomagnets each ($N = 20 \times 20 \times 2 = 800$). These samples are large enough to suppress even minute finite-size effects (edge effects), as has been checked by comparison with calculations for larger arrays. The dynamics is obtained by kinetic Monte Carlo simulations, accompanied by standard Monte Carlo calculations\cite{Binder79,Metropolis87} (see Appendix~\ref{sec:Monte}).

\section{Discussion of results}
\label{sec:discussion}
In the following, we focus on samples with vertical displacements $\delta z$ of $0$ and $0.27\,a$, as well as on temperatures $T\approx\unit[1]{K}$ and $\unit[300]{K}$ (room temperature).

\subsection{Magnetic ground state}
\label{sec:magnetic-ground-state}
\begin{figure}
 \centering
 \includegraphics[width = 0.89\columnwidth]{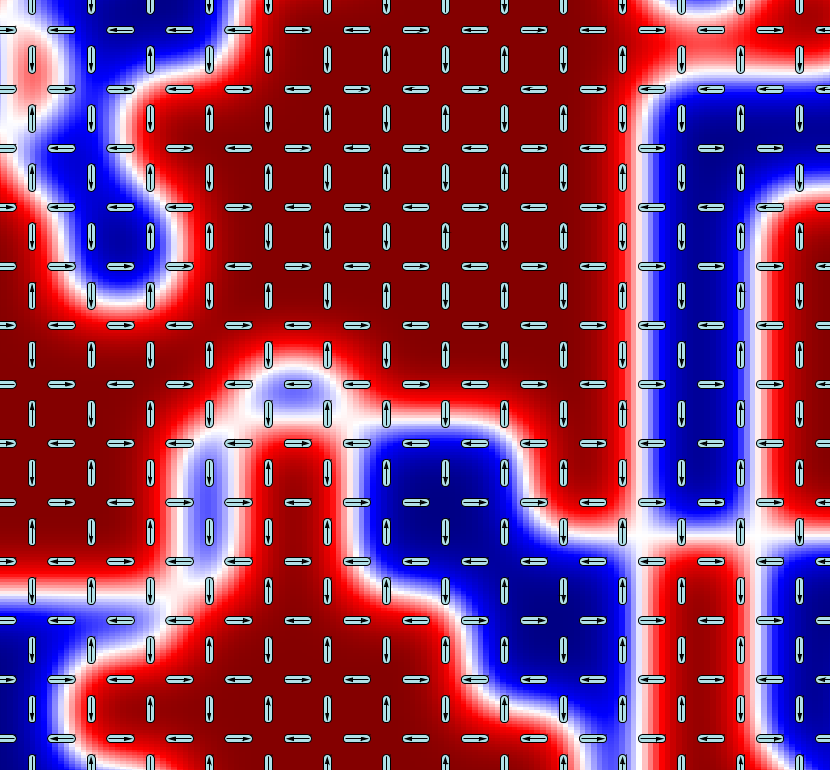}
 \caption{(Color online) Magnetic ground state in square-lattice spin ice for a vertical displacement of $\delta z = 0.27\,a$ at $T=\unit[0]{K}$. All six `2in2out' vertices are equally likely. The background color indicates  `2in2outOp' (red) and `2in2outAd' (blue) domains.}
 \label{fig:domain}
\end{figure}

For \amend{a small finite} temperature \amend{of $T\approx\unit[1]{K}$}, we find a ground state in agreement with the ice rule (Figure~\ref{fig:domain}); hence, irrespectively of $\delta z$ one has $\eta_{0} = \unit[100]{\%}$. A closer inspection shows that `2in2outOp' vertices dominate for $\delta z = 0$, in agreement with earlier work (e.\,g., Ref.~\onlinecite{Mol10}). Upon increasing $\delta z$, the number of `2in2outAd' vertices grows. Especially at $\delta z = 0.27\,a$, all six `2in2out' vertices are equally likely; this is explained by the energy barrier between the `2in2outAd' and `2in2outOp' vertices which vanishes for this particular vertical displacement.

The system tends to form `2in2outOp' and `2in2outAd' domains, where the shape of the domains depends on the numerical `cooling-down' procedure used to obtain the global, highly degenerate free-energy minimum (Appendix~\ref{sec:Monte}).  For $\delta z > 0.27\,a$, `2in2outAd' vertices prevail. Elevated temperatures lead to changes of size and to propagation of domains.

\subsection{Thermal string excitations}
\label{sec:thermal-excitations}

\subsubsection{Thermal activation and switching rates}
\amend{We now show that the square-lattice dipolar arrays are thermally active at room temperature and that the rate of spin reversals depends significantly on the vertical displacement $\delta z$}. Thermal activity at $\unit[300]{K}$ cannot be ruled out \textit{per se} because the maximum nearest-neighbor interaction energy $E_{\mathrm{1NN}}$ of $\unit[9.2]{meV}$ is less than the thermal energy $k_{\mathrm{B}} T \approx \unit[25]{meV}$; cf.\ Fig.~\ref{fig:energies}.

According to the implementation of the kinetic Monte Carlo method (Ref.~\onlinecite{Kratzer09} and Appendix~\ref{sec:Monte}), the rate $\tau^{-1}$ of spin reversals  scales exponentially with temperature and the energy barrier, since $\tau^{-1}$ follows an Arrhenius form. The barrier height depends on the initial and final configurational energies $E_{\mathrm{i}}$ and $E_{\mathrm{f}}$  and is assumed linear\cite{Fichthorn00}: $\Delta E = E_{0} + \nicefrac{1}{2} (E_{\mathrm{f}} - E_{\mathrm{i}})$, where $E_{0}$ is an empirical parameter taken from Ref.~\onlinecite{Farhan13}. 

\begin{figure}
 \centering
 \includegraphics[width = 0.99\columnwidth]{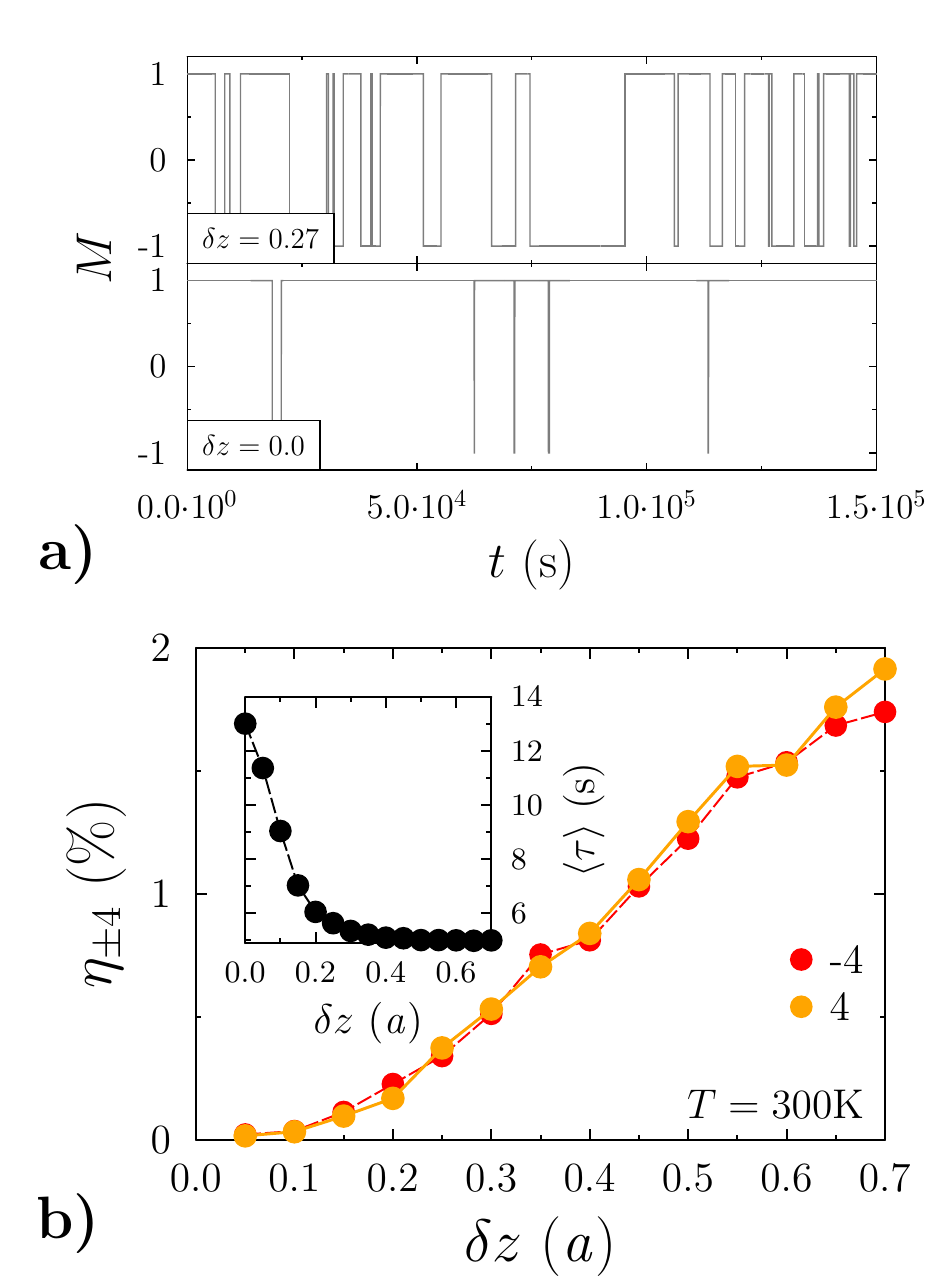}
 \caption{(Color online) Thermal activation of square-lattice spin ice. (a) Representative time sequences of reversals of a selected nanomagnet, obtained from kinetic Monte Carlo simulations, are shown for vertical displacements $\delta z = 0.27\,a$ and $0.0$ (indicated in each panel) at room temperature $T = \unit[300]{K}$. $M = \pm 1$ characterizes the orientation of the selected magnetic moment. (b) Fractions of charges $\eta_{\pm 4}$ versus vertical displacement $\delta z$ ($a$ lattice constant). The inset displays the average rest time $\langle \tau \rangle$ between consecutive reversals \amend{of the entire sample}.}
 \label{fig:thermal}
\end{figure}

\amend{Thermal activation is addressed by the duration---or rest time---between reversal of nanoislands. Figure~\ref{fig:thermal}a shows two representative sequences of magnetization reversal of one selected nanoisland; these could be measured by a local probe. Obviously, the reversal rate is larger for $\delta z = 0.27\,a$ as compared to that for $\delta z = 0$; in other words, the rest time becomes smaller with increasing $\delta z$. For the sequences shown, we obtain average rest times of $\unit[4.4\cdot10^3]{s}$ and $\unit[1.3\cdot 10^4]{s}$ for $\delta z = 0.27\,a$ and $0$, respectively, at $T = \unit[300]{K}$.}

\amend{Similar to the rest time of a single island, one can record the rest time of an entire sample. This duration is defined as the time between reversals of any nanoislands in the array. For arrays with $800$ islands at $T = \unit[300]{K}$, we obtain average rest times of $\unit[5.4]{s}$ and $\unit[12.4]{s}$ for $\delta z = 0.27\,a$ and $0$, respectively. These values are smaller than that of a single island; they scale inversely with the number of islands in the sample; more precisely, they are about $1/800$ of the single-island rest time. Because of the abovementioned Arrhenius behavior, rest times decrease significantly with temperature: for $T = \unit[420]{K}$, as has been  applied in Ref.~\onlinecite{Farhan13}, our simulations yield durations of the order of a few milliseconds.}

\amend{We point out that the rest times should not be confused with the residence time defined in Ref.~\onlinecite{Farhan13}. The residence time is defined as the duration between the reversal of the flux chirality of a plaquette ($\unit[26]{s}$ for the hexagonal rings studied in Ref.~\onlinecite{Farhan13}). Such a definition is somewhat problematic for a square lattice because its plaquettes must not show flux closure.}

For zero vertical displacement $\delta z = 0$, the ground state `2in2outOp' nodes result in closed loops for the plaquettes; this can be viewed as energy-minimizing  `flux closures'. This is not the case for $\delta z = 0.27\,a$, for which there are  `2in2outAd' nodes in addition (Fig.~\ref{fig:string}). This loss of flux closure is explained by the increased degeneracy of the `2in2out' nodes and a considerable number of nodes with charge $Q = \pm 2$; see top row in Fig.~\ref{fig:string}.

\subsubsection{Number of string excitations}
\begin{figure*}
 \centering
 \includegraphics[width = 1.0\textwidth]{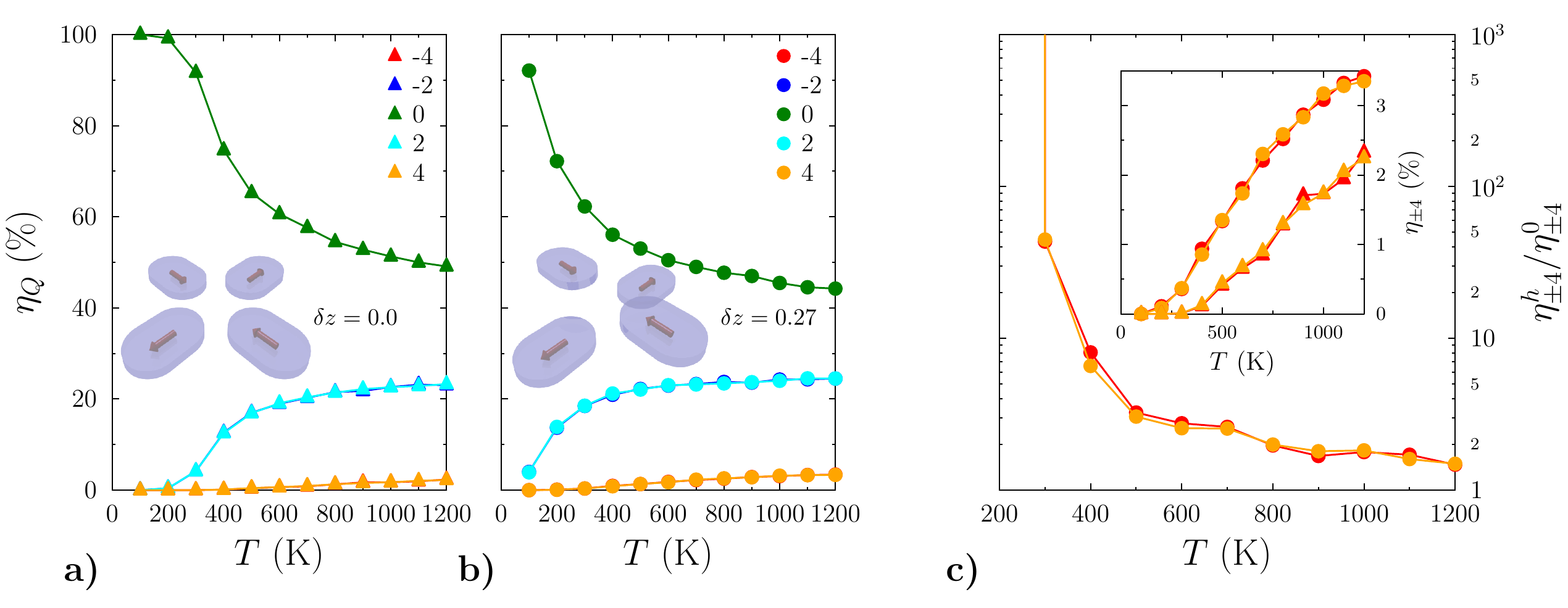}
 \caption{(Color) Magnetic charges $Q$ in square-lattice dipolar arrays. The fractions $\eta_{Q}$ of charges $Q = -4, \ldots, +4$ are shown for lattices with vertical displacement $\delta z = 0$ (a) and $\delta z = 0.27\,a$ (b) versus temperature $T$. Because $\eta_{Q} = \eta_{-Q}$ on average, data for negative charges are covered by those for positive $Q$. (c) Ratios $\eta_{\pm 4}(0.27\,a) / \eta_{\pm 4}(0)$ on a logarithmic scale versus temperature $T$. Monopole-charge fractions $\eta_{\pm4}$ for $\delta z = 0$ (triangles) and $\delta z = 0.27\,a$ (circles) [same data as in (a) and (b)] are given in the inset. Data are obtained by kinetic Monte Carlo simulations.}
 \label{fig:charge}
\end{figure*}

A finite temperature below the critical temperature of the nanoislands leads to thermal excitations with nonzero charge\cite{Kapaklis12} (Fig.~\ref{fig:charge}; note that $\langle \eta_{Q} \rangle = \langle \eta_{-Q} \rangle$): the larger $|Q|$, the smaller is $\langle \eta_{\pm Q} \rangle$. In particular, $\langle \eta_{\pm 4} \rangle$ is less than $\unit[2.3]{\%}$ for samples with $\delta z = 0$ at elevated temperatures; at room temperature it is extremely small.

A closer inspection reveals, however, that $\langle \eta_{\pm 4} \rangle$ is strongly enhanced for $\delta z = 0.27\,a$ as compared to samples with $\delta z = 0$ (Fig.~\ref{fig:charge}c). More precisely, there are about $\unit[3.5]{\%}$ quasi-monopoles with $|Q| = 4$ in the sample at $\unit[1200]{K}$. Compared with the fraction of $\unit[2.3]{\%}$ for $\delta z = 0$, this increase may be regarded insignificant. However at room temperature, we find an enhancement by a factor as large as $43$ [inset in Fig.~\ref{fig:charge}(c)]. Vertical displacement is, therefore, a means to enhance the number of excitations; their number may be sufficiently large to allow investigations of ensembles of string excitations\cite{Gliga13}.

So far, we considered the fractions of nonzero charges in a sample. That string excitations are present is evident from a snapshot of a kinetic Monte Carlo simulation (Fig.~\ref{fig:string}). While a large part of the sample shows a ground-state configuration, there is also a single string excitation: a path of ferromagnetically aligned nanomagnets (green nanomagnets in Fig.~\ref{fig:string}) connects a quasi-monopole of charge $-4$ (indicated by the blue circle, with `4out' arrangement) with a quasi-monopole of charge $+4$ (red circle, with `4in' arrangement).

\subsubsection{Spatial correlation of string excitations}
The spatial distribution of nodes with opposite charges is analyzed by means of the charge-correlation function 
\begin{align}
 S_{\nu\mu}\left(\left|\delta\vec{r}\right|\right) \equiv \langle Q^{\nu}_{\vec{r}} \, Q^{\mu}_{\vec{r}+\delta \vec{r}}\rangle_{\vec{r}}
 \label{eq:correlation}
\end{align}
which defines the probability of simultaneously finding a charge $Q = \mu$ at position $\vec{r} + \delta\vec{r}$ and a charge $Q = \nu$ at position $\vec{r}$. The average is over all nodes in the sample, thus $S_{\nu\mu} = S_{\mu\nu}$. 

\begin{figure}
 \centering
 \includegraphics[width = 0.45\textwidth]{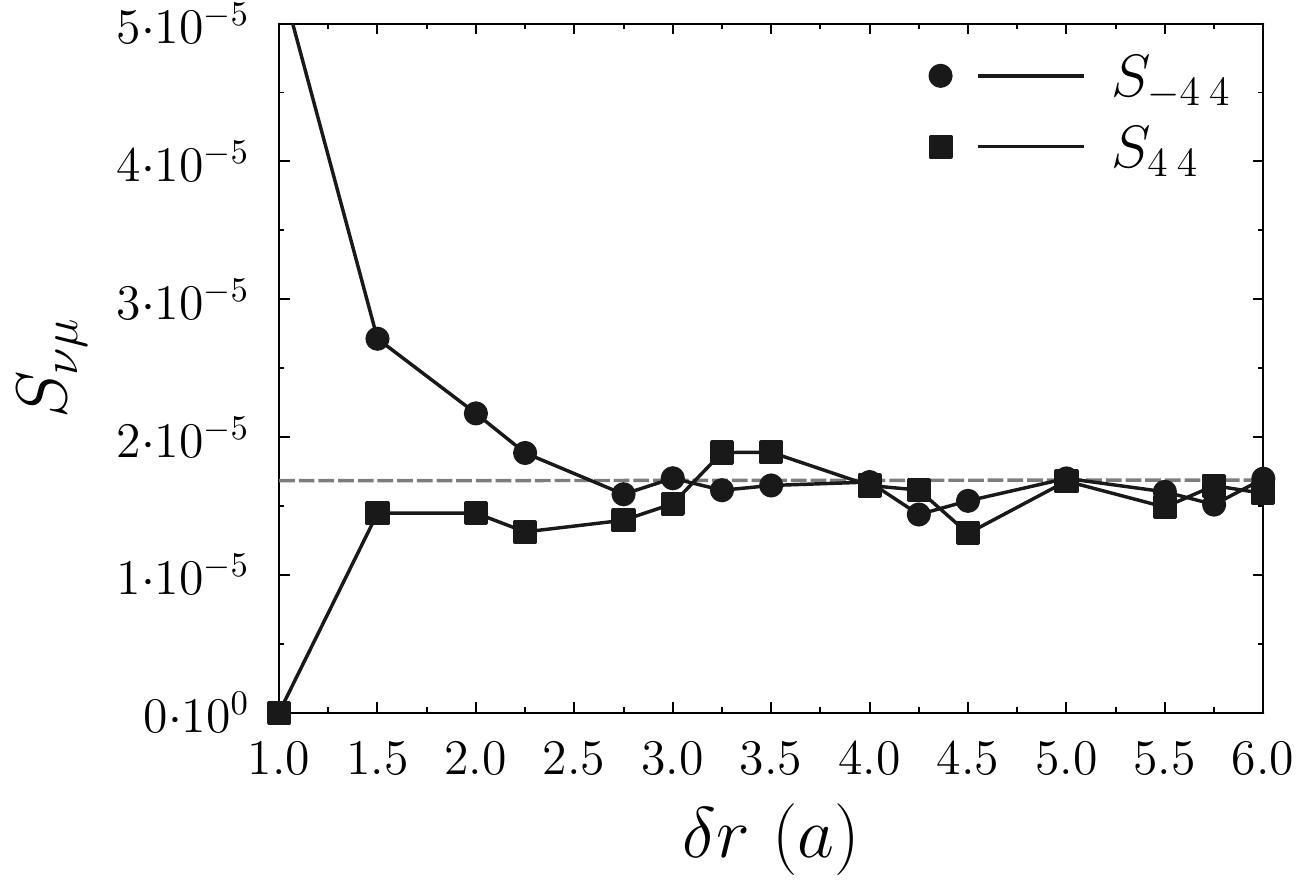}
 \caption{Charge-correlation function $S_{\nu\mu}$ [eq.~(\ref{eq:correlation})] versus distance $\delta r$ in square-lattice dipolar arrays at $T=\unit[300]{K}$. Data are shown for $S_{-4\, 4}$ (circles) as well as for $S_{4\, 4}$  (open squares). $\delta r$ is in units of the lattice constant $a$. The dotted line indicate the saturation level for large distances.}
 \label{fig:correlation}
\end{figure}

It turns out that $S_{-4\, 4}$ is nonzero within the first four shells of neighbor nanomagnets (circles in Fig.~\ref{fig:correlation}). According to the free-energy minimization, these pairs prefer to arrange with the shortest possible distance. Pairs of nodes with identical charges $Q = 4$ cannot show up as nearest neighbors because of the lattice geometry (one nanomagnet would be shared among a pair). $S_{4\, 4}$ shows no clear indication for short distances $\delta r$ rather than a uniform distribution (open squares in Fig.~\ref{fig:correlation}).

\section{Concluding remarks}
Square-lattice dipolar arrays prove suitable for studying thermal string excitations in artificial spin ice. By varying the vertical displacement of rows and columns---for example done by microstructuring techniques---one can produce samples with a prescribed temperature dependence of the string-excitation density. The thermal stability (mean average time) can be chosen to match the time resolution of the experimental probing technique.

Future investigations may focus on the effect of defects in the dipolar arrays (e.\,g., missing islands) or on the formation of domains.

\acknowledgments
We thank Marin Alexe for fruitful discussions.

\appendix

\section{Interaction energies}
\label{sec:energies}
The Heisenberg-type exchange is neglected in the calculations, owing to the fact that the nanomagnets are isolated from each other. Thus, the dominant coupling mechanism comes from the dipole-dipole interaction.  The total interaction energy then reads
\begin{align}
 E & = -\sum_{i \not= j} \vec{m}_{i} \cdot \boldsymbol{\mathsf{Q}}_{ij} \cdot \vec{m}_{j}.
\end{align}
Here, the magnetization density $\vec{m}_{i} = \nicefrac{\vec{M}_{i}}{\Omega_{i}}$, where $\Omega_{i}$ is the volume of the $i$-th island, is assumed homogeneous.

Expressing a position $\vec{r}$ within the $i$-th nanomagnet by $\vec{r} \equiv \vec{R}_{i} + \vec{u}_{i}$, where  $\vec{u}_{i}$ runs over its volume $\Omega_{i}$, the elements of the dipole-dipole tensor $\boldsymbol{\mathsf{Q}}_{ij}$ are
\begin{align}
 Q_{ij}^{\mu\nu}
 & =\frac{\mu_0}{8\pi} 
 \int_{\Omega_{i}}
 \int_{\Omega_j}
  \frac{3 r^{\mu}_{ij} r^{\nu}_{ij} - \delta_{\mu\nu} {\vec{r}_{ij}}^{\,2}}{\vec{r}_{ij}^{\,5}}
  \,\mathrm{d}\vec{u}_{j}
  \,\mathrm{d}\vec{u}_{i},
  \label{eq:Q}
\end{align}
with $\mu, \nu = x, y, z$. Here, $\vec{r}_{ij} \equiv \vec{u}_{i} + \vec{R}_{i} - \vec{u}_{j} - \vec{R}_{j}$ and $\mu_0$ is the vacuum permeability.

Besides analytical calculations, we use numerical integration schemes for the evaluation of the dipole-dipole tensor because these allow to treat arbitrarily shaped nanomagnets. For the present study, the integrals in eq.~(\ref{eq:Q}) are performed using a Gauss-Legendre quadrature with 32 supporting points in each spatial direction. As a consequence of taking into account the experimental geometry of Ref.~\onlinecite{Farhan13}, the energy cross-over $E_{\mathrm{1NN}} = E_{\mathrm{2NN}}$ (Fig.~\ref{fig:energies}) occurs at a vertical displacement $\delta z$ that is different from those calculated with a shape approximation for the nanomagnets; for example $0.419\,a$ for needles\cite{Moeller06} and $0.444\,a$ for points\cite{Mol10}.

It turns out that the first and the second nearest neighbors provide the relevant contributions to the interaction energy; more precisely, $E_{\mathrm{3NN}} = 0.045 E_{\mathrm{1NN}}$ and $E_{\mathrm{4NN}} = 0.07 E_{\mathrm{1NN}}$ for $\delta z = 0.27\,a$, with $E_{\mathrm{1NN}}$ being the first-nearest neighbor interaction energy. Interactions of second- and third-nearest neighbors do not depend on $\delta z$.

Lithographic techniques allow to produce nanomagnets with a specific shape. The chosen shape has evidently impact on the interaction energies, although the lattice spacing may be unaltered. Here, we briefly compare the interaction energies of two types with rectangular shape. Type 1 is strictly rectangular with an aspect ratio of $2.76$ (as in Ref.~\onlinecite{Farhan13}), Type 2 is a rounded island with the same area as type 1, compose of a rectangular with an aspect ratio of $1.98$ and a circles with radius $\unit[85]{nm}$.

Having computed the set $\{ \boldsymbol{\mathsf{Q}}_{ij} \}$ of dipole tensors, we proceed with statistical methods that work on a discrete set in space (lattice of nanomagnets) and in the spin degrees of freedom (orientations of the nanomagnets' magnetizations).

\section{Monte Carlo and kinetic Monte Carlo calculations}
\label{sec:Monte}
To simulate the ground state as well as the dynamics of the artificial spin ice, Monte Carlo\cite{Binder97b,Boettcher12} and kinetic Monte Carlo calculations have been performed. Both methods are implemented in the \textsc{cahmd} computer code\cite{cahmd,Boettcher10}.

A Monte Carlo method tries to find a global minimum of the free energy at a given temperature $T$ by successively reversing the island spins $\vec{M}_{i}$. Using the Metropolis algorithm\cite{Metropolis53}, the reoriented state (final state) is accepted, if the energy difference $\Delta E = E_{\mathrm{f}} - E_{\mathrm{i}}$ between the initial and the final state  is negative or if the Boltzmann factor $\exp{\left(-\nicefrac{\Delta E}{k_{\mathrm{B}} T}\right)}$ is larger than a uniformly distributed random number $p \in [0, 1]$. $k_{\mathrm{B}}$ is the Boltzmann constant.

In a kinetic Monte Carlo method, the reorientation rate $r_{i}$ for each spin $\vec{M}_{i}$ in the lattice follows an Arrhenius law,
\begin{align}
 r_{i} & = \rho_{0} \,\exp\left(\frac{-\Delta E_{i}}{k_{\mathrm{B}} T}\right).
 \label{eq:rate}
\end{align}
$\Delta E_{i}$ is the site-dependent energy barrier, while $\rho_{0}$ is a fundamental rate fitted to experiment. 

At each kinetic Monte Carlo step, cumulative rates $\Gamma_{i} \equiv \sum_{j = 1}^{i} r_{j}$ are calculated for $i = 1, \ldots, N$ ($N$ number of nanomagnets). Then, the magnetization of the $i$-th island is reversed, if $\Gamma_{i-1} \leq p \cdot \Gamma_{N} < \Gamma_{i}$, with the random number $p$ uniformly distributed in $[0, 1]$ and $\Gamma_{0} \equiv 0$. The rest time $\tau$, that is the duration between two successive reversals \amend{in the entire sample}, is $\tau = \Gamma_{N}^{-1}\ln(\nicefrac{1}{p^\prime})$ ($p^\prime$ uniformly-distributed random number).

The energy barrier $\Delta E_{i}$ in eq.~(\ref{eq:rate}) is given by the dipolar energy and depends on the initial and the final state of the entire system. In the present work, it is assumed linear\cite{Fichthorn00}: $\Delta E_{i} \equiv E_{0} + \nicefrac{1}{2} (E_{\mathrm{f}} - E_{\mathrm{i}})$. The larger $E_{0}$, the smaller are the rates and the larger are the rest times.  $\rho_{0}$ and $E_{0}$ are empirical parameters and taken from Ref.~\onlinecite{Farhan13} ($E_{0} = \unit[0.925]{eV}$, $\rho_{0} = \unit[10^{-12}]{s^{-1}}$). Both our standard and kinetic Monte Carlo approaches reproduce well the correlation functions and the switching rates for the hexagonal rings studied by Farhan \textit{et al.} (Ref.~\onlinecite{Farhan13}).

The energy barrier depends on the dipole energy variation including the vertical displacement of the islands which increases the reorientation rate. In the picture of a Stoner-Wohlfarth double well potential, $E_{0}$ is determined by the magnetic anisotropy as well as by the inter-atomic magnetic exchange mechanisms\cite{Boettcher11}.

For both standard and kinetic Monte Carlo simulations, an initial `cooling down', starting at $T = \unit[5000]{K}$ and approaching the chosen temperature of the simulation in 10\,000 steps, has been performed to come close to the global free-energy minimum. A typical kinetic Monte Carlo simulation comprises at least 100\,000 steps, with magnetic configurations saved to disk in intervals of 1000 steps. Average rest times $\langle \tau \rangle$ have been computed using all steps, while average charge fractions $\langle \eta_{Q} \rangle$ are calculated from 100 samples.

\section{Residual entropy}
\label{sec:entropy}
Following Pauling\cite{Pauling35}, a pyrochlore lattice contains $Z = (\nicefrac{3}{2})^{\nicefrac{N}{2}}$ microstates for $N$ spins, leading to the entropy per spin of $S = \nicefrac{k_{\mathrm{B}}}{N} \ln Z\approx 0.2 k_{\mathrm{B}}$ (the factor of $2$ comes from the two possible spin orientations). Considering a step-by-step build-up of a finite, vertically displaced spin-ice cluster from the top-left to the bottom-right corner, the ground state of a node with the center $C_i$ is dominated by the configuration of its top and left node in the two adjacent islands. Depending on the relative alignment of the island spin coming from the top-node and the left-node, one obtains four possible states at node $i$. Neglecting rim effects, the number of states ends up with $Z = (\nicefrac{3}{2})^{\nicefrac{N}{2}}$ and the same residual entropy per spin for zero temperature as predicted by Pauling\cite{Pauling35} for water ice. For $\delta z = 0$, however, the entropy per spin is zero, corroborating a `quasi-ice' character of such a system.

\bibliographystyle{apsrev}

\end{document}